\documentclass{article}
\usepackage[main,final]{neurips_2026}
\usepackage[utf8]{inputenc}
\usepackage[T1]{fontenc}
\usepackage{hyperref}
\usepackage{orcidlink}
\usepackage{url}
\usepackage{booktabs}
\usepackage{amsfonts,amsmath}
\usepackage{microtype}
\usepackage{graphicx}
\usepackage{xcolor}
\definecolor{linknavy}{HTML}{1F3A5F}  %
\hypersetup{colorlinks=true, citecolor=linknavy, linkcolor=linknavy, urlcolor=linknavy}

\title{Online Identification Can Replace\\a Floor-Plan Prior on Unseen Buildings}

\author{%
  Elias T.-H. Shieh\,\orcidlink{0009-0006-2120-8830}\\
  Siebel School of Computing and Data Science\\
  University of Illinois Urbana-Champaign\\
  \texttt{tshieh2@illinois.edu}
  \And
  Chun Fu\thanks{Corresponding author.}\\
  Department of Bioenvironmental Systems Engineering\\
  National Taiwan University\\
  \texttt{chunfu@ntu.edu.tw}
}

\begin{document}
\maketitle

\begin{abstract}
The Smart Buildings Challenge asks for one HVAC control policy that runs office buildings and weather it has never seen, from the building's metadata and its sensor streams alone.
Our controller sees no data from the target building before it runs and is not tuned per building; its few global constants were set on development scenarios.
Each zone has a first-order grey-box model that starts from coefficients computed from the floor plan and is updated online by recursive least squares, and a planner and a closed-form mapping turn the models into plant commands.
On two sets of buildings and weather that we generated, a validation set and a stress set with real-year weather, it scores 0.145 and 0.159 times the starter kit's PID baseline, and 0.095 times on a holdout built after all versions were fixed and scored once.
It improves on every building, also against our two implementations of ASHRAE Guideline 36: per set it scores 0.60 to 0.71 times a basic version and 0.83 to 0.96 times one partly tuned on the public leaderboard, which wins one of 54 scenarios; with a 5\,\textdegree C frost target for empty zones on both, the margin over the tuned version shrinks to 1.4\% on the stress set, where one of six buildings is 0.2\% worse.
In an ablation on a reduced-order model, removing both the floor-plan prior and online identification costs 28\%, 20\% and 42\% on the validation, stress and holdout sets, far more than the two effects combined: either source of building-specific information recovers most of what is lost without both. On one office floor in a second simulator, BOPTEST, this does not carry over: with identification on, the two priors give similar results, but the floor-plan prior alone, with constants fitted on the competition's simulator, is the least stable arm.
These simulation results motivate testing floor-plan-initialised adaptive control in buildings that have drawings but no operating data.
\end{abstract}

\section{Introduction}
\label{sec:intro}
The Smart Buildings Challenge \citep{goldfeder2026challenge} tests transfer: one policy runs 30 hidden scenarios that combine office buildings and weather it has not seen, and it receives only the building's metadata and its sensor streams.
Model predictive control needs a model of each building, and learned controllers move that cost into training (\S\ref{sec:related}).
We take the opposite route to training a general policy: the controller learns nothing offline about the target building.
Each zone's first-order model starts from coefficients computed from the zone map and is updated online by recursive least squares; a planner and a closed-form mapping turn the models into device commands (\S\ref{sec:method}).

We ask what transfers to an unseen building.
Our contributions:
\begin{itemize}
  \item A controller with no per-building tuning that beats the PID baseline and two Guideline 36 controllers in every building of two generated sets and a holdout (Figure~\ref{fig:scenarios}, Table~\ref{tab:main}); with a 5\,\textdegree C frost target for both (ours soft, its floor hard), its margin over the tuned one shrinks to 1.4\% on the stress set.
  \item An ablation of the building-specific parts of the model: the floor-plan prior and online identification are largely redundant on the competition's simulator (reduced-order model; on one BOPTEST case the prior's constants do not carry over), and freezing identification costs almost twice as much on the stress set with cold real-year weather as on the validation set, and least on the holdout (\S\ref{sec:results}).
  \item An evaluation protocol for a benchmark with known accounting issues: a stock and a billing-corrected simulator, decisions per building, and a stress set that caught a tuning the validation set had passed (\S\ref{sec:eval}, \S\ref{sec:results}).
  \item A measurement of how cold the score lets empty buildings get and what a frost target costs (\S\ref{sec:analysis}).
\end{itemize}

\section{Related work}
\label{sec:related}

\paragraph{The benchmark.}
The Smart Buildings Challenge \citep{goldfeder2026challenge} is built on the Smart Buildings Control Suite \citep{goldfeder2024suite}.
Whether a policy trained on some buildings carries over to others has become a benchmark question of its own \citep{taboga2026building2building}.
An earlier HVAC-control competition let participants tune by rerunning the scored scenarios, which no real building allows \citep{walnum2026adrenalin}; this one hides its buildings and weather.

\paragraph{Model-based control and its modelling cost.}
Model predictive control can save energy in buildings without giving up comfort, but each building needs its own control-oriented model, and developing that model is the main bottleneck in practice \citep{oldewurtel2012mpc,drgona2020allyouneed}.
Online identification of building models by recursive least squares \citep{grubinger2016gotl} or Kalman filters \citep{radecki2015online} is established, and offset-free MPC adds a disturbance estimate for what the model misses; \citet[\S7]{drgona2020allyouneed} review both.

\paragraph{Transfer to unseen buildings.}
Transfer-learning approaches reach unseen buildings with models trained on other buildings' data \citep{grubinger2016gotl,hossain2020bnn,raisch2025gentl,raisch2026neural,raisch2026veryexciting}, one combining an offline source model with an online recursive-least-squares model of the target \citep{grubinger2016gotl}; ours carries over only a few global constants.
The trade-off between prior knowledge and data in estimating building parameters is known \citep{pathak2019prior,raillon2019bayesian}, and transferred models and policies have been tested in closed loop on unseen buildings \citep{grubinger2016gotl,taboga2026building2building,raisch2026neural,raisch2026veryexciting}.
Building information also initialises adaptively identified models in \citet{wan2023semi} (BIM geometry and materials; closed loop in one simulated zone) and \citet{bagle2026adaptive} (zone area and energy standard; prediction in one real room); we test how a zone-map prior and online identification interact, in a factorial closed-loop ablation on unseen buildings.

\section{Problem and metric}
\label{sec:problem}

\paragraph{Task.}
A submission is one policy class.
For each scenario it is constructed once with the building's metadata (zone map, observation and action indices, action ranges, comfort window and the scheduled occupied hours) and then called every 5 minutes with the current observation for 30 days (8{,}641 rewarded steps; Appendix~\ref{app:repro}).
The 30 hidden scenarios combine buildings and weather that the policy has not seen; nothing identifies the building or the weather station.
Observations are the sensor and actuator readings of every VAV box and air handler (AHU), the outside air temperature, the building's occupant count and the time (Appendix~\ref{app:obs}).
Observations look backwards: zone temperatures lag the plant by one step (a host ruling, forum \href{https://www.kaggle.com/competitions/neurips-2026-smart-buildings/discussion/749945}{749945}),\footnote{Forum $n$: \nolinkurl{kaggle.com/competitions/neurips-2026-smart-buildings/discussion/}$n$; PR $n$: \nolinkurl{github.com/smart-buildings-2026/starter-kit/pull/}$n$.} and the occupant count averages the previous step, so it can show occupants for one step after everyone has left.

\paragraph{Reward and score.}
Each step is scored
\begin{equation}
  r_t = 0.6\,p_t - 0.2\,e_t - 0.2\,c_t,
  \label{eq:reward}
\end{equation}
where $p_t \le 0$ is the normalised productivity regret of zones outside the comfort window, $e_t$ the normalised energy cost and $c_t$ the normalised carbon emission.
The regret is zero whenever nobody is in the building, so unoccupied hours cost only energy and carbon.
A scenario scores $|\bar{r}|$, its absolute mean reward, and the leaderboard averages the 30 scenarios equally; lower is better.

\paragraph{Two accounting issues in the simulator.}
Both change the score without a physical cause and were reported publicly by other participants.
First, boiler gas is billed against the hot-water setpoint while the reheat coils draw on the actual tank temperature, so a policy that holds the setpoint below the tank temperature receives unbilled heat (starter-kit \href{https://github.com/smart-buildings-2026/starter-kit/pull/2}{PR~\#2}, forum \href{https://www.kaggle.com/competitions/neurips-2026-smart-buildings/discussion/749531}{749531}).
The hosts aim to run the final verification on a corrected simulator and ask controllers to stay physically consistent in the meantime (forum \href{https://www.kaggle.com/competitions/neurips-2026-smart-buildings/discussion/751998}{751998}).
Our policy holds the setpoint at its maximum at every step (\S\ref{sec:plant}), and we score every candidate on both the stock and the corrected simulator (\S\ref{sec:eval}).
Second, each AHU's return air is the floor-area-weighted mean of its zones although the zones exchange heat in proportion to their airflow, so uneven airflow creates or destroys heat (starter-kit \href{https://github.com/smart-buildings-2026/starter-kit/pull/3}{PR~\#3}, forum \href{https://www.kaggle.com/competitions/neurips-2026-smart-buildings/discussion/753804}{753804}); Appendix~\ref{app:return} measures what our policy receives.

\paragraph{What the score does not see.}
Because regret is zero in unoccupied hours, nothing in \eqref{eq:reward} penalises how cold an empty building gets; the simulator also models neither freezing nor humidity (forums \href{https://www.kaggle.com/competitions/neurips-2026-smart-buildings/discussion/753664}{753664}, \href{https://www.kaggle.com/competitions/neurips-2026-smart-buildings/discussion/752030}{752030}).
The hosts invite a description of frost protection and its cost; we report one in \S\ref{sec:analysis}.

\section{Method}
\label{sec:method}

\begin{figure}[!tbh]
  \centering
  \includegraphics[width=\linewidth]{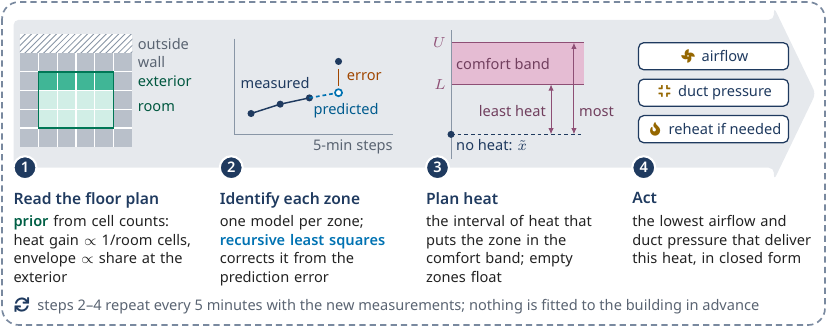}
  \caption{The controller, drawn schematically; nothing is fitted to the building in advance. Each zone's model starts from the floor-plan prior (step 1, with the room's cells at the exterior dark; \S\ref{sec:priors}) and is corrected by recursive least squares (step 2, \S\ref{sec:model}); the planner gives each zone the interval of heat that reaches the comfort band (step 3, \S\ref{sec:planner}); a closed-form mapping turns it into device commands (step 4, \S\ref{sec:plant}).}
  \label{fig:overview}
\end{figure}

The controller is a single Python file using numpy, plus \texttt{scipy.ndimage} to label rooms in the zone map.
It is rebuilt for every scenario and uses only the metadata (indices, comfort window, schedule hours, zone map) and the observation stream.
Each 5-minute step runs three layers (Figure~\ref{fig:overview}): a per-zone model updated online (\S\ref{sec:model}), a planner that gives each zone an approximate admissible heat interval (\S\ref{sec:planner}), and a closed-form mapping from zone heat to device commands (\S\ref{sec:plant}).
\emph{Notation:} zone $i = 1,\dots,Z$; $\mathcal{Z}_a$ the zones of air handler (AHU) $a$; $y_{k,i}$ the observed and $x_{k,i}$ the modelled zone temperature at step $k$ (K); $T^{\mathrm{o}}_k$ the outside air (mean of the AHU sensors); $c_k = 1$ on weekdays within the scheduled hours, else $0$; $[T_{\mathrm{lo}}, T_{\mathrm{hi}}]$ the comfort window; $[z]_{l}^{u}$ clips $z$ to $[l, u]$.

\subsection{Per-zone grey-box model}
\label{sec:model}

Each zone has a linear first-order model with coefficients $\theta_i \in \mathbb{R}^6$:
\begin{equation}
  x_{k+1,i} - x_{k,i} = \theta_i^{\top}\phi_{k,i},\qquad
  \phi_{k,i} = \bigl[\,T^{\mathrm{o}}_k - x_{k,i},\; Q_{k,i},\; m_{k,i} - x_{k,i},\; \bar{x}_k - x_{k,i},\; c_k,\; 1\,\bigr]^{\top},
  \label{eq:model}
\end{equation}
with terms for the envelope, the heat $Q_{k,i}$ (kW, negative for cooling) delivered by the zone's VAV box, a slow state $m_{k,i}$ standing in for the slab, the unweighted building mean $\bar{x}_k$, an offset during scheduled hours, and a constant.
The slow state is an exponential average, $m_{k+1,i} = m_{k,i} + (x_{k+1,i} - m_{k,i})/24$.
$Q$ is computed from the policy's own commands, $Q_{k,i} = C_{k,i}(T^{\mathrm{d}}_{k,i} - x_{k,i})/1000$, with $C_{k,i}$ the commanded air capacity rate (W/K) and $T^{\mathrm{d}}_{k,i}$ the commanded discharge temperature (\S\ref{sec:plant}).
The coefficients are updated every step by per-zone recursive least squares (RLS) with forgetting factor $\lambda = 0.9995$:
\begin{equation}
  g = \frac{P\phi}{\lambda + \phi^{\top}P\phi},\quad
  e = \bigl[\Delta y - \phi^{\top}\theta\bigr]_{-0.5}^{0.5},\quad
  \theta \leftarrow \theta + g\,e,\quad
  P \leftarrow \bigl(P - g\,\phi^{\top}P\bigr)/\lambda .
  \label{eq:rls}
\end{equation}
Because observed temperatures lag the plant by one step, the policy advances them one model step with its last command before planning; this estimate $\hat{x}_k$ and $m_k$ start all planning (Appendix~\ref{app:lag}). In \eqref{eq:rls}, $\Delta y$ is the observed one-step temperature change.
\label{sec:dist}A per-zone bias $d_i$, an exponential average of the one-step residual with a 36-minute time constant, covers disturbances that change within a day, faster than the week-long RLS memory (Appendix~\ref{app:dist}).

\subsection{Priors from the zone map}
\label{sec:priors}

\begin{figure}[!tbh]
  \centering
  \includegraphics{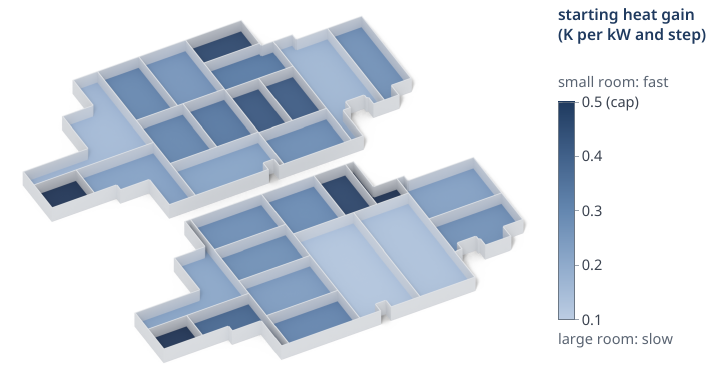}%
  \caption{The floor-plan prior as the controller reads it: the zone map of validation-set building\_10002 (two floors), each room coloured by its starting heat-gain coefficient $\theta^{0}_{2,i}$ of \eqref{eq:prior} with $s = 1$ (cold weather; rooms of at most 50 cells keep $\theta^{\mathrm{g}}$). RLS then corrects these values.}
  \label{fig:floorplan}
\end{figure}

A single global prior $\theta^{\mathrm{g}}$ (medians of offline fits; Appendix~\ref{app:details}) ignores that zones differ in size and exposure.
The metadata's zone map is the simulator's floor plan (interior space 0, walls 1, exterior 2), whose rooms are the 4-connected components of interior space; on the 42 buildings we checked (2076 rooms), the policy's labels match the simulator's \texttt{room\_$N$} cell for cell.
The policy counts each room's cells $n_i$ and exterior-contact cells $e_i$ (Appendix~\ref{app:rooms}).
The heat gain and envelope coefficients then start from
\begin{equation}
  \theta^{0}_{2,i} = \bigl[\, s \cdot 1000 / n_i \,\bigr]_{0.01}^{0.5},\qquad
  \theta^{0}_{1,i} = \bigl[\, 1.1\, e_i / n_i \,\bigr]_{0}^{0.05},
  \label{eq:prior}
\end{equation}
with the scale $s$ set once from the first outside-air reading: 1 in cold weather and 0.35 in warm weather, where the full prior overestimated the heat gain (Appendix~\ref{app:scale}).
A zone's response to delivered heat scales inversely with its air volume and its coupling to outside with the exterior share of its floor: in per-zone least-squares fits to logs of all 30 scenarios of the development set (six development buildings, \S\ref{sec:eval}), the fitted heat gains correlate with $1/n_i$ at 0.83 to 0.996 (median 0.96) and the envelope coefficients with $e_i/n_i$ at 0.91 to 0.995 (median 0.98).
Rooms of at most 50 cells, or all rooms if the map cannot be parsed, keep $\theta^{\mathrm{g}}$.
Figure~\ref{fig:floorplan} shows the geometry-dependent part of the starting model: on one validation building, small rooms start with a large heat gain per kW of delivered heat and large rooms with a small one.

\subsection{Planner}
\label{sec:planner}

A step is \emph{occupied} on a weekday within the scheduled hours or while occupants are present (holiday and early-departure rules in Appendix~\ref{app:occ}).

\emph{Admissible heat interval.}
The target band is $[L, U] = [T_{\mathrm{lo}} + 0.28,\, T_{\mathrm{hi}} - 0.26]$.
The model is rolled forward $H = 2$ steps from $(\hat{x}_k, m_k)$ with the current outside temperature, once with $Q = 0$ (free response $x^{\mathrm{f}}_H$) and once with $Q = 1$\,kW in every zone ($x^{1}_H$).
With gain $G_i = \max(x^{1}_{H,i} - x^{\mathrm{f}}_{H,i}, 10^{-4})$ and $\tilde{x}_i = x^{\mathrm{f}}_{H,i} + H d_i$, the planner uses the interval
\begin{equation}
  \bigl[q^{\mathrm{lo}}_i,\, q^{\mathrm{hi}}_i\bigr] = \bigl[(L - \tilde{x}_i)/G_i,\; (U - \tilde{x}_i)/G_i\bigr]
  \label{eq:interval}
\end{equation}
as the heat that puts the predicted temperature at step $H$ in the band.
This is an approximation, not a guarantee, because the building-mean term couples the zones (Appendix~\ref{app:coupling}).
Occupied steps use this interval; unoccupied steps use $(-\infty, \infty)$, so zones float.

\emph{Optimal start.}
Before each weekday start, the model is rolled forward to the start once without and once with heat; a zone heats at full capacity now only if starting a little later, with 75\% of its modelled capacity, would miss the band (Appendix~\ref{app:start}).

\emph{Night flush and evening trim.} Before a weekday that is expected to need cooling, AHUs left off by the planner are run on cool outside air to bank cooling in the building mass, and on weekdays the AHUs and the pump are switched off for the step before the evening hour and the step at it, when the reward already counts the building as empty (Appendix~\ref{app:flush}).

\subsection{Plant mapping}
\label{sec:plant}

The plant layer converts each zone's interval to commands in closed form with the simulator's device equations: the lowest airflow that delivers the zone's heat, the lowest duct pressure that reaches it, and reheat only where the supply air falls short (Appendix~\ref{app:plantmap} and \ref{app:plant}).

The hot-water supply setpoint is held at its maximum, 350\,K, at every step.
The simulator bills boiler gas against this setpoint while the reheat coils use the actual tank temperature, so a setpoint below the tank temperature gives unbilled reheat (\S\ref{sec:problem}); ours never goes below it.
The policy takes about 2.7\,ms per step (at most 28\,s per scenario on the validation and stress sets).

A CMA-ES search \citep{hansen2016cma} on short windows of 12 development-set scenarios, selected on 12 further scenarios run on the reduced model (\S\ref{sec:eval}), set ten constants; it also changed two RLS constants, which we reverted after the stress set (\S\ref{sec:constants}).

\section{Evaluation protocol}
\label{sec:eval}

\paragraph{Scenario sets.}
The official training set has six buildings and 30 scenarios.
To evaluate on unseen buildings and weather, we generated sets with the starter kit's generators, sampling building sizes near those of the six official buildings rather than the generator's defaults (Appendix~\ref{app:repro} lists them):
\begin{itemize}
  \item The \textbf{validation set} (val18p): 6 new buildings $\times$ 3 of the 18 official weather strips that the training set does not use (18 scenarios). These strips are milder than the real-year months of the stress set; the hottest Phoenix summers are all in the training set.
  \item The \textbf{stress set} (hard18x): 6 further buildings, three of them four storeys high, $\times$ 3 strips of real-year airport observations (NOAA ISD-Lite), so that every building gets one hot and one cold month (18 scenarios). It is harder: the PID baseline scores 0.101 on it against 0.034 on the validation set and 0.064 on the public leaderboard.
  \item A \textbf{development set} of 30 scenarios on six other generated buildings (dev30p) and 12 reduced-model selection scenarios for tuning; the official training set (\textbf{train30}) only for regression checks.
\end{itemize}

\paragraph{Formal scores.}
A formal score (our ``gate'') runs every scenario to the end in the official simulator with the submitted policy file, on the stock simulator and on the boiler-corrected copy (Appendix~\ref{app:gate}).

The ablations in \S\ref{sec:results} use a reduced-order model of the simulator's thermal grid (ROM; Appendix~\ref{app:rom}), whose scenario scores for our policy differ from the formal ones by at most 0.0075\% on the 36 scenarios where we have both; we mark every number that comes from it.

\paragraph{Comparisons.}
Two policies are compared by their score ratio on the same set; seeds are fixed, so spreads are across scenarios and buildings, not reruns. Scenarios sharing a building are not independent, so the primary evidence is how many of the six buildings improve (6 of 6; a two-sided sign test gives $p = 0.031$, nominal on the two sets that selected versions) and the ratio's range with one building left out.
The validation set and, from the hand-set version (v1n) on, the stress set decided which versions we kept (Appendix~\ref{app:selection}); neither is a clean held-out test, so \S\ref{sec:results} adds one.

\section{Results}
\label{sec:results}

\begin{figure}[!tbh]
  \centering
  \includegraphics[width=\linewidth]{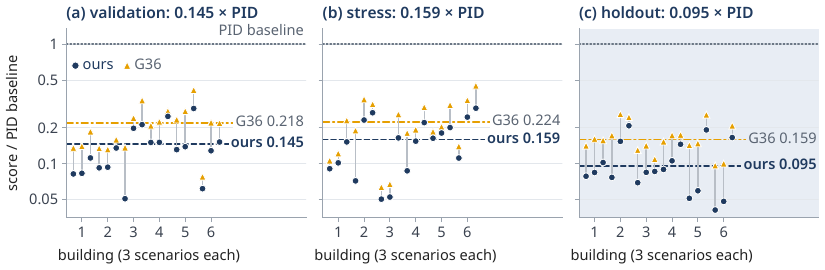}
  \caption{Every scenario improves on the PID baseline and a basic Guideline 36 baseline (G36; the tuned G36-t is closer, Table~\ref{tab:main}), including a holdout scored once after all versions were fixed (formal scores, stock simulator; ratio to PID, log scale; lower is better). Each panel is one set of six unseen buildings with three weather scenarios each: (a) official weather strips and (b) the stress set, real-year weather with a hot and a cold month per building (\S\ref{sec:eval}); (c) the holdout (Appendix~\ref{app:comparisons}). One mark per scenario: navy dot ours, orange triangle G36; dashed lines and the values at their right: each controller's ratio over all 18. (a) and (b) informed which versions we kept; (c) did not.}
  \label{fig:scenarios}
\end{figure}

\paragraph{Main result.}
Table~\ref{tab:main} compares our submitted policy (v1r) with the starter kit's PID baseline (\texttt{pid\_water}; \citealp{goldfeder2026challenge}), two rule-based controllers following ASHRAE Guideline 36 \citep{ashrae2021g36}, our first model-based version (v0c) and the hand-set version before the tuned constants.
The submitted policy scores 0.145 times the PID baseline on the validation set and 0.159 times on the stress set, and every scenario improves on both sets (the worst is 0.29; Figure~\ref{fig:scenarios}).
A holdout (hold18) built after all versions were fixed and scored once per policy, without selecting anything (six new buildings, each with three 2023 strips from six U.S. airports no other set uses; Appendix~\ref{app:comparisons}) gives 0.095 times (0.0040 against 0.0421), better in all 18 scenarios (95\% bootstrap interval over buildings 0.075--0.119; Figure~\ref{fig:scenarios}c).
Both Guideline 36 controllers were built for this benchmark by one of us (C.F.; Appendix~\ref{app:g36}) from the guideline's sequences, with no building model or learned weights: a basic version (G36) and the team's competition entry (G36-t), tuned in part on the public leaderboard. The submitted policy scores 0.667, 0.709 and 0.598 times G36 and 0.899, 0.959 and 0.833 times G36-t on the three sets, better in every building and every scenario but one (stress set, 1.010 times G36-t). Unlike ours, both keep a floor for empty zones (12\,\textdegree C; G36-t down to $-3$\,\textdegree C in deep cold), which the score does not reward. With a 5\,\textdegree C frost target added (\S\ref{sec:analysis}), the submitted policy scores 0.918, 1.008 and 0.840 times G36-t and still improves on G36 in every scenario of every set (at most 0.88 times). Raising G36-t's deep-cold floor to the same 5\,\textdegree C costs it 1.006, 1.023 and 1.000; against this frost-matched G36-t the ratios are 0.912, 0.986 and 0.840, with five cold stress-set scenarios (at most 1.032) and one stress-set building (1.002) worse; our target is soft and its floors are hard (Appendix~\ref{app:g36}).
With the billing corrected, none of our versions scores worse (at most 0.5\% better) and G36 and G36-t change by at most 0.4\%, while the PID baseline is 5.6\% worse: it benefits from the gap.
Public scores were excluded from the selection rules of our model-based versions; on that different scenario set, the submitted policy scores 0.00996, 0.156 times the PID baseline.
Our team's leaderboard entries also include submissions from a separate line that uses the simulator's boiler billing gap; no result here comes from it or, apart from one diagnostic in Appendix~\ref{app:unseen}, from any policy using the gap, and each public score is that policy's own submission.
Other results use the submitted policy; Appendix~\ref{app:later} gives our latest version (v1y).

\begin{table}[t]
  \centering
  \caption{Mean $|\bar r|$ over scenarios (lower is better): formal on the official simulator (stock) and its billing-corrected copy (boilerfix); Kaggle public leaderboard (other scenarios; $^\dagger$partly tuned on it). --: not run; frost target: a soft 5\,\textdegree C planning target for unoccupied zones (\S\ref{sec:analysis}); deep-cold floor: G36-t's unoccupied floor in deep cold raised from $-3$ to 5\,\textdegree C (Appendix~\ref{app:g36}).}
  \label{tab:main}
  \small
  \begin{tabular}{lccccc}
    \toprule
    & \multicolumn{2}{c}{validation set} & stress set & holdout & \\
    policy & stock & boilerfix & stock & stock & public \\
    \midrule
    PID baseline (starter kit) & 0.034134 & 0.036033 & 0.10132 & 0.042057 & 0.06404 \\
    Guideline 36, basic (G36) & 0.007430 & 0.007431 & 0.022678 & 0.006700 & 0.01385 \\
    Guideline 36, tuned (G36-t) & 0.005513 & 0.005493 & 0.016779 & 0.004805 & 0.01044$^\dagger$ \\
    \quad with 5\,\textdegree C deep-cold floor & 0.005548 & -- & 0.017160 & 0.004804 & -- \\
    first model-based (v0c) & 0.005674 & 0.005670 & 0.017416 & -- & 0.01095 \\
    hand-set constants (v1n) & 0.005115 & 0.005115 & 0.016532 & -- & 0.01026 \\
    submitted, analysed here (v1r) & 0.004956 & 0.004956 & 0.016088 & 0.004004 & 0.00996 \\
    \quad with 5\,\textdegree C frost target & 0.005060 & -- & 0.016915 & 0.004034 & -- \\
    latest on 7 November (v1y) & 0.004821 & 0.004798 & 0.015503 & -- & 0.00961 \\
    \bottomrule
  \end{tabular}
\end{table}

\begin{figure}[t]
  \centering
  \includegraphics[width=\linewidth]{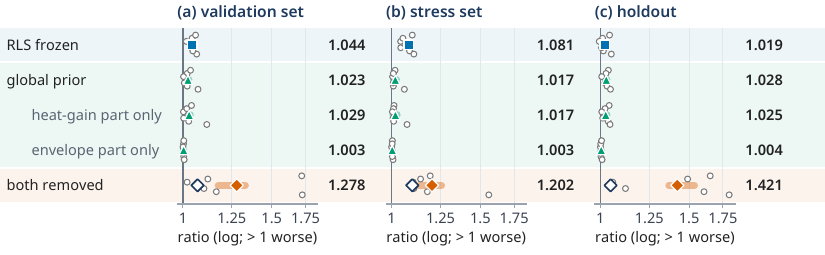}
  \caption{Ablation of the model layer on the three sets (reduced model, ratio to the submitted policy; $>1$ is worse). Filled marker and number: ratio over all 18 scenarios; bar: its range when one building is left out (hidden where shorter than the marker; values in Appendix~\ref{app:interaction}); circles: the six buildings (overlapping where nearly equal). Open diamond: the product of the RLS-frozen and global-prior rows, the cost without interaction (1.069, 1.099 and 1.048).}
  \label{fig:ablation}
\end{figure}

\paragraph{What transfers: prior and identification.}
Either source of building-specific information, the floor plan before any data or the first days of data, recovers most of what is lost without both (Figure~\ref{fig:ablation}; reduced model, ROM, all 18 scenarios of each set, \S\ref{sec:eval}), and the holdout, run after the result on the other two sets was known, repeats it.
The arms freeze the RLS update at the zone-map prior, replace that prior by the global $\theta^{\mathrm{g}}$ (all of it, or only its heat-gain or envelope part) with RLS on, or do both.
Freezing costs almost twice as much on the stress set as on the validation set; the stress set's buildings differ as well as its weather.
Removing both the prior and the update costs 28\% on the validation set, 20\% on the stress set and 42\% on the holdout, against 6.9\%, 9.9\% and 4.8\% if the two effects did not interact.
The losses differ in kind: almost all of the frozen arm's loss is comfort regret (its energy and carbon even fall slightly), almost all of the global-prior arm's loss is energy and carbon, and the arm with both removed loses in both terms.
On all three sets, the prior's effect comes mostly from its heat-gain part (from room area). The prior carries the first days and the transients, identification the weeks after (Figure~\ref{fig:time} in Appendix~\ref{app:time}). All arms keep the fast disturbance estimate.

\paragraph{A second simulator.}
We repeated the ablation on one BOPTEST case \citep{blum2021boptest} (a five-zone office floor and its four official two-week heating and cooling scenarios; Appendix~\ref{app:boptest}, Table~\ref{tab:boptest}).
The model, prior, RLS and planner are the submitted policy's, with the competition's constants unchanged; only the interface to the plant is rewritten (control points, observation timing, measured heat input, ventilation and schedule).
BOPTEST reports several indicators and no single score, so we do not rank the arms.
All four arms improve on the case's own controller in thermal discomfort, energy, cost and emissions in every scenario (energy 0.63--0.98 times), while CO$_2$ discomfort is higher in the two heating scenarios (272--4{,}767\,ppm\,h per zone against zero) and for the global-prior arm in typical cooling.
With identification on, the two priors are again close: the full controller has 5--24\% less thermal discomfort than the global-prior arm and uses 0.1--6.2\% more or less energy.
The rest does not carry over: removing both is not the worst arm but has the lowest or equal discomfort and the highest or second-highest energy.
The floor-plan prior with identification frozen is the least stable: frozen, the perimeter zones' envelope coupling stays at the prior's cap of 0.05 per step, where identification settles on 0.012--0.018 in heating, so the model overstates heat loss; in the typical heating scenario this arm uses 39\% more energy than the full controller with 2.5 times its discomfort.

\paragraph{The stress set catches what the validation set does not.}
The constants tuned by CMA-ES improved the validation set by a factor of 0.9755 with all 18 scenarios better, yet on the stress set they scored 1.041 times the hand-set version, the net loss coming from one building (building\_14006, 1.31; two others were also worse; Figure~\ref{fig:stress}a in Appendix~\ref{app:selection}) and within it from one extreme-cold scenario (\S\ref{sec:constants}).
Reverting the two RLS constants fixed it; without the stress set we would have kept the tuned constants.
For ten pairs of consecutive submitted versions, local ratios predicted public-leaderboard ratios within 1\% (0.3\% for the last six; Figure~\ref{fig:stress}b).

\section{What the score leaves out for empty buildings}
\label{sec:analysis}

With no regret outside occupied hours, our policy lets empty zones float.
In the coldest stress-set scenario 27\% of zone-steps are below 5\,\textdegree C and 19\% below 0\,\textdegree C, the coldest zone reaches $-30.3$\,\textdegree C, and the coldest occupied zone is at 9.8\,\textdegree C (ROM; Figures~\ref{fig:cold} and~\ref{fig:coldnight} and Table~\ref{tab:unseen} in Appendix~\ref{app:frost}).
A soft frost target (an extra lower end of each zone's heat interval) costs a factor of 1.077 (ROM) on the six cold stress-set scenarios for a 5\,\textdegree C target; it is a planning target, not a guarantee (Appendix~\ref{app:frost}). In formal scores over all 18 scenarios of each set, the 5\,\textdegree C target costs 1.021, 1.051 and 1.007 on the validation, stress and holdout sets (Table~\ref{tab:main}).

\section{Limitations and conclusion}
\label{sec:limits}\label{sec:conclusion}
The validation and stress sets come from the starter kit's generators, so they test transfer within the benchmark's family of buildings, not to real buildings; the only test outside it is one BOPTEST case with four scenarios (\S\ref{sec:results}).
With six buildings per set, building-level intervals are descriptive, and the ablation and per-scenario frost numbers are ROM results; the ROM was checked against the official simulator for the submitted policy only.
The ablation's global-prior arm replaces the prior with fixed coefficients, not with no prior, and we do not identify why the two sources are redundant.
Redrawing the validation set's occupancy, as the final verification will (forum \href{https://www.kaggle.com/competitions/neurips-2026-smart-buildings/discussion/752678}{752678}), moves its ratio to the hand-set version from 0.969 to 0.968 (Appendix~\ref{app:comparisons}); the stress set and ablations were not rerun.
On 10th November 2026 the best public score of the policies in this paper was 0.00961 (v1y), 2.0 times the leading team's 0.00470; without other teams' policies we do not attribute this gap to a cause.

Started from its floor plan and corrected by its own data, a controller untrained on the target building scores 0.095--0.159 times the PID baseline on unseen buildings, including a holdout scored once; on the competition's simulator, either source recovers most of what is lost without both, while on one BOPTEST case the floor-plan constants did not carry over.
On the validation and stress sets the two sources work at different times: the prior carries the first days and the transients, identification the weeks after.
This matters where modelling each building is the cost of model-based control: a new or retrofitted building has drawings before it has operating data, and on the competition's simulator the drawings were enough to start from and the first weeks of the building's own data enough to correct them.
In practice they could be a BIM model and the commissioning period; the next step is a real building, with a prior not fitted to its simulator.

\begin{ack}
We used AI assistants throughout this work. Anthropic's Claude and OpenAI's Codex wrote and ran most of the experiment code and drafted most of the text under our direction. Claude and OpenAI's GPT models also reviewed the drafts and checked the reported numbers against the experiment logs, and Google's Gemini was used for a readability check. We chose the research questions, decided which results to report and revised the text, and we take full responsibility for all of the content. Chun Fu's laboratory at National Taiwan University provided the Claude subscription used in this work and supported travel to NeurIPS 2026. The work received no other funding.
\end{ack}

\bibliographystyle{plainnat}
\bibliography{refs}

\appendix

\section{Method details}
\label{app:details}

\paragraph{Model constants.}
The global prior is $\theta^{\mathrm{g}} = (0.018,\, 0.115,\, 0.31,\, 0.05,\, 0.3,\, 0.08)$ per 5-minute step, the medians of offline least-squares fits on logged \texttt{pid\_water} runs of one development building in three climates; the prior covariance is $P_0 = \mathrm{diag}(10^{-5}, 10^{-3}, 10^{-3}, 10^{-3}, 10^{-2}, 10^{-3})$.
After each RLS update the coefficients are projected onto $\theta_1 \in [0, 0.05]$ (envelope), $\theta_2 \in [0.002, 0.5]$ (heat gain), $\theta_3 \in [0, 0.8]$ (slab) and $\theta_4 \in [0, 0.3]$ (building mean); the schedule and constant terms are free.
The slow state $m$ is initialised to the first observation.
The zone-map prior falls back to $\theta^{\mathrm{g}}$ for every zone if \texttt{scipy} is missing, the map is not two-dimensional, or a VAV number falls outside the labelled rooms.

\paragraph{Tuned constants.}
CMA-ES values (hand-set v1n values in brackets): band margins 0.283 and 0.260\,K (0.4, 0.4); $H = 2$ (3); $\tau_d = 7.28$ steps (9); $\kappa = 0.752$ (0.7); optimal-start lead $\ell = 2$ (3); holiday after 0.979\,h (1.0); night-flush target $T_{\mathrm{lo}} + 0.777$\,K (0.9), window 44 steps (48), minimum outside-air margin 3.30\,K (4.0).
Kept at the hand-set values: $\lambda = 0.9995$ and the 24-step slab constant (\S\ref{sec:constants}).

\paragraph{Night-flush rules.}
A day counts as a cooling day if the previous day had at least 12 occupied steps in which a running AHU had a supply-air setpoint below 304\,K, or if the outside air exceeded 295\,K at any step yesterday or so far today.
On such a day, while unoccupied and $0 < S_k \le 44$, with $S_k$ the number of steps until the next weekday morning start (the last 3\,h\,40\,min before it), an AHU that the planner left off flushes if, on observed temperatures, $T^{\mathrm{o}}_a \le T^{\mathrm{ret}}_a - 3.3$\,K, its warmest zone is at least $T_{\mathrm{lo}} + 1.08$\,K and its coldest at least $T_{\mathrm{lo}} + 0.6$\,K.
It then runs with the supply-air setpoint at 305\,K, 200\,Pa static pressure, reheat valves closed, and recirculation $r = [(289 - T^{\mathrm{o}}_a)/(T^{\mathrm{ret}}_a - T^{\mathrm{o}}_a)]_0^{0.7}$ if $T^{\mathrm{o}}_a < 289$\,K (which tempers the mix to 289\,K) and $r = 0$ otherwise.
Dampers are fully open on zones above $T_{\mathrm{lo}} + 0.98$\,K and at the 10\% floor elsewhere.
The resulting $C_i = \rho c_p F^{\max}_i \delta_i \sqrt{200/2000}$ and $T^{\mathrm{d}}_i = T^{\mathrm{mix}}_a$ replace that step's command record.

\subsection{Plant mapping and capacity equations}
\label{app:plant}

Symbols follow \S\ref{sec:method}; $\hat{x}_i$ is the estimated zone temperature and $\rho c_p = 1.2 \times 1006$\,J\,m$^{-3}$\,K$^{-1}$.
For AHU $a$, $T^{\mathrm{ret}}_a$ is the mean estimated temperature of its zones and $T^{\mathrm{mix}}_a = r\,T^{\mathrm{ret}}_a + (1-r)\,T^{\mathrm{o}}_a$ the mixed air for recirculation share $r \le 0.7$.
$F^{\max}_i$ is the zone's observed supply-airflow setpoint and $C^{\mathrm{full}}_i = \rho c_p F^{\max}_i$ its air capacity rate at full damper and 2000\,Pa.
$T^{\mathrm{w}}$ is the observed supply-water temperature floored at 310\,K, so it always exceeds the discharge cap $T^{\mathrm{d}}_{\max} = 308.15$\,K.
The policy models a reheat valve's water capacity rate as $C^{\mathrm{w}} = v\,W\sqrt{\Delta p}$ for opening $v$ and pump head $\Delta p$ [bar], with $W = \rho_{\mathrm{w}} c_{\mathrm{w}} K_v / 3600$, $K_v = 0.108$\,m$^3$/h, and $\bar{C}^{\mathrm{w}} = W\sqrt{20}$ at full valve and maximum head.
$K_v$ is the \texttt{vav\_reheat\_water\_flowrate} that both scenario generators write and that every configuration we used carries (the 30 training scenarios, the sample evaluation scenario, dev30p, val18p and hard18x).
The simulator gives valve $i$ the share $v_i K_v / \sum_j v_j K_v$ of the pump flow $\sum_j v_j K_v \sqrt{\Delta p / (1 + (\sum_j v_j K_v)^2 R)}$; no configuration sets the header resistance $R$, so $R = 0$ and each valve passes exactly $v_i K_v \sqrt{\Delta p}$, as modelled.

\paragraph{Capacities used by the planner.}
Heating assumes $r = 0.7$ and the supply-air setpoint at its maximum, $T^{\mathrm{s}} = \min(T^{\mathrm{mix}}_a, 305)$.
With the valve fully open, $C \ge \bar{C}^{\mathrm{w}}$ and the discharge below the cap, heat into the zone is $C(T^{\mathrm{s}} - \hat{x}) + \bar{C}^{\mathrm{w}}(T^{\mathrm{w}} - T^{\mathrm{s}})$, which falls with airflow when $T^{\mathrm{s}} < \hat{x}$; the most heat is delivered at the airflow where the fully open valve brings the discharge air exactly to the cap:
\begin{gather}
  C^{*}_i = \min\Bigl(C^{\mathrm{full}}_i,\; \frac{\bar{C}^{\mathrm{w}}(T^{\mathrm{w}} - T^{\mathrm{s}})}{\max(T^{\mathrm{d}}_{\max} - T^{\mathrm{s}}, 0.5)}\Bigr),\qquad
  \bar{Q}^{\mathrm{h}}_i = \frac{C^{*}_i (T^{\mathrm{d}}_i - \hat{x}_i)^{+}}{1000},\\
  T^{\mathrm{d}}_i = \min\Bigl(T^{\mathrm{d}}_{\max},\; T^{\mathrm{s}} + \min\bigl(1, \bar{C}^{\mathrm{w}}/C^{*}_i\bigr)(T^{\mathrm{w}} - T^{\mathrm{s}})\Bigr).
\end{gather}
Cooling capacity is full airflow at the coldest supply air, $\bar{Q}^{\mathrm{c}}_i = C^{\mathrm{full}}_i(\hat{x}_i - \min(285, T^{\mathrm{mix}}_a))^{+}/1000$ with $T^{\mathrm{mix}}_a$ at $r = 0.7$.

\paragraph{Airflow and static pressure.}
Given an AHU's supply temperature $T^{\mathrm{s}}$, each zone's desired air capacity rate is
\begin{equation}
  C^{\mathrm{want}}_i =
  \begin{cases}
    \min\bigl(C^{\mathrm{full}}_i,\; -1000\,q^{\mathrm{hi}}_i / (\hat{x}_i - T^{\mathrm{s}})\bigr) & \text{if } q^{\mathrm{hi}}_i < 0,\ \hat{x}_i > T^{\mathrm{s}} + 0.2,\\[2pt]
    \min\Bigl(C^{\mathrm{full}}_i,\; \dfrac{1000\,q^{\mathrm{lo}}_i}{\max(T^{\mathrm{d}}_{\max} - \hat{x}_i, 0.5)},\; C^{\mathrm{kink}}\Bigr) & \text{if } q^{\mathrm{lo}}_i > 0,\\[4pt]
    0 & \text{otherwise}.
  \end{cases}
\end{equation}
(the first case that holds applies), where $C^{\mathrm{kink}} = \bar{C}^{\mathrm{w}}(T^{\mathrm{w}} - T^{\mathrm{s}})^{+}/\max(T^{\mathrm{d}}_{\max} - T^{\mathrm{s}}, 0.5)$ is the flow at which the fully open valve heats the supply air exactly to the cap, as in $C^{*}_i$.
The static pressure is $P = [\,2000 \max_{i \in \mathcal{Z}_a} (C^{\mathrm{want}}_i / C^{\mathrm{full}}_i)^2\,]_{200}^{2000}$\,Pa, the delivered rate is $C_i = [C^{\mathrm{want}}_i]_{0.1\hat{C}_i}^{\hat{C}_i}$ with $\hat{C}_i = C^{\mathrm{full}}_i\sqrt{P / 2000}$, and the damper command is $\delta_i = C_i/\hat{C}_i$.

\paragraph{Reheat and pump.}
If $C_i(T^{\mathrm{s}} - \hat{x}_i)/1000 < q^{\mathrm{lo}}_i$, the zone needs $T^{\mathrm{req}}_i = \min(\hat{x}_i + 1000\,q^{\mathrm{lo}}_i / C_i,\; T^{\mathrm{d}}_{\max})$, effectiveness $\varepsilon_i = [(T^{\mathrm{req}}_i - T^{\mathrm{s}})/(T^{\mathrm{w}} - T^{\mathrm{s}})]_0^1$, water capacity rate $C^{\mathrm{w}}_i = \varepsilon_i C_i$, and discharge temperature $T^{\mathrm{d}}_i = T^{\mathrm{s}} + \varepsilon_i(T^{\mathrm{w}} - T^{\mathrm{s}})$; otherwise $C^{\mathrm{w}}_i = 0$ and $T^{\mathrm{d}}_i = T^{\mathrm{s}}$.
Over all zones of the building, the pump head is $\Delta p = [(1.05 \max_i C^{\mathrm{w}}_i / W)^2]_1^{20}$\,bar and each valve opens to $v_i = [C^{\mathrm{w}}_i/(W\sqrt{\Delta p})]_0^1$; with no reheat anywhere, $\Delta p = 0$, all valves close and the pump is off.
The commanded $C_i$ and $T^{\mathrm{d}}_i$ are stored as the inputs of the next model update (\S\ref{sec:model}).

\paragraph{Supply-air setpoint and recirculation under cooling.}
For an AHU with some $q^{\mathrm{hi}}_i < 0$, each candidate $T^{\mathrm{sp}} \in \{285, 287, 289, 291, 293\}$\,K is evaluated with $r = 0.7$, or, if $T^{\mathrm{o}}_a < T^{\mathrm{ret}}_a - 1$, with the economiser share that brings the mix to the setpoint, $r = [1 - (T^{\mathrm{ret}}_a - T^{\mathrm{sp}})/(T^{\mathrm{ret}}_a - T^{\mathrm{o}}_a)]_0^{0.7}$.
With the zone flows above and total airflow $V = \sum_i C_i/(\rho c_p)$, the cost is
\begin{equation}
  J = 2\Bigl(\frac{V P\,(2 - r)}{0.9} + \frac{\rho c_p V (T^{\mathrm{mix}}_a - T^{\mathrm{s}})^{+}}{3.5}\Bigr) + \sum_{i} C_i (T^{\mathrm{d}}_i - T^{\mathrm{s}})^{+} + 10^{6} \sum_{i} \Bigl(\frac{C_i (T^{\mathrm{d}}_i - \hat{x}_i)}{1000} - q^{\mathrm{hi}}_i\Bigr)^{+},
  \label{eq:satcost}
\end{equation}
fan power (fan efficiency 0.9, with the factor $2 - r$ adding the exhaust fan for the outside-air share), chiller power (COP 3.5) and reheat power, with electricity weighted twice gas (in the score the ratio per watt lies between 1.81 and 2.35, Appendix~\ref{app:tou}) and a penalty when the zones receive more heat than $q^{\mathrm{hi}}$ allows.
The setpoint with the lowest $J$ is kept.
All continuous commands are mapped linearly onto $[-1, 1]$ and placed by the metadata's action indices; an AHU whose zones need no action is switched off with its dampers and valves at their minimum commands.
Capacity estimates use the observed supply-water temperature, while the hot-water setpoint is always commanded at 350\,K.

\section{Further details}
\label{app:more}

\subsection{Fast disturbance estimate}\label{app:dist}
With $\lambda = 0.9995$ the RLS memory is about $1/(1-\lambda) = 2000$ steps, roughly a week, too slow for disturbances that change within a day.
A per-zone bias on the one-step residual covers them; at each accepted update, before \eqref{eq:rls},
\begin{equation}
  d_i \leftarrow d_i + \tfrac{1}{\tau_d}\bigl[\Delta y_i - \theta_i^{\top}\phi_i - d_i\bigr]_{-0.5}^{0.5},
  \label{eq:dist}
\end{equation}
an exponential average of the residual with a time constant of $\tau_d = 7.3$ steps (36 minutes).
It is active only while the building counts as occupied or the next occupied period is within 2\,h, and zero otherwise.
It enters the lag compensation and the planner's free response, but not RLS; the optimal-start rollouts start from $\hat{x}_k$, which already includes $d$ once, but do not add it at each step.
Every arm of the ablation in \S\ref{sec:results} keeps it; removing it costs 4.5\% on val18p and 5.6\% on hard18x (formal gate, v1n).

\subsection{Coupling error of the heat interval}\label{app:coupling}
The interval \eqref{eq:interval} is an approximation, not a guarantee.
Model \eqref{eq:model} is affine in $Q$, but the building-mean term couples the zones, giving an $H$-step response $x_H = x^{\mathrm{f}}_H + K Q$ with non-diagonal $K$.
Here $G = K\mathbf{1}$ is the response to equal heat in every zone, and $K Q = G \odot Q$ holds only when all zones receive the same heat or $K$ is diagonal.
Since $\theta_4 \le 0.3$ and the mean averages over all zones, the off-diagonal entry $K_{ij}$ grows by about $\theta_{4,i}\,\theta_{2,j}/Z$ per step, and the error is largest when one zone is heated hard while its neighbours float.
For the two-step horizon of v1r and v1x (Appendix~\ref{app:later}) the error does not depend on the state, $e_i = \bigl((K - \mathrm{diag}(G))Q\bigr)_i = \theta_{4,i}\bigl(\tfrac{1}{Z}\sum_j \theta_{2,j} Q_j - Q_i\,\tfrac{1}{Z}\sum_j \theta_{2,j}\bigr)$.
We evaluated it on v1x's logs of the 30 dev30p scenarios with the delivered heat of each step and coefficients fitted offline to the same logs ($\theta_4$ clipped to the policy's range), not the policy's online estimates: over 5.1 million occupied zone-steps the median of $|e_i|$ is 0.002\,K, the 99th percentile 0.085\,K (2.8\% of the 3\,K comfort band) and the maximum 0.41\,K (14\%).

\subsection{Night flush and evening trim}\label{app:flush}
\emph{Night flush.}
In the last 44 steps (3\,h\,40\,min) before a weekday that is expected to need cooling, any AHU left off by the planner is run on outside air when the outside air temperature is at least 3.3\,K below its return air and its zones are warm enough to bank cooling in the building mass.
The supply-air setpoint is held at its maximum, so the supply equals the mixed air and the chiller stays off; pressure is at the 200\,Pa minimum and reheat is closed.
The flush airflow is written into the command record so that RLS learns from it; the triggering rules are in Appendix~\ref{app:details}.

\emph{Evening trim.}
On weekdays the AHUs and the pump are switched off for the step before the metadata's evening hour and the step at it (18:55 and 19:00 in all our scenarios), when the reward already counts the building as empty.
This relies on one assumption about the configuration: that the latest departure of the occupancy model equals the metadata's evening hour, which holds in all 166 scenarios we generated from the official templates; if a hidden scenario lets occupants stay later, the trim switches the AHUs off while someone is present.
Because the observed occupant count averages the previous step (\S\ref{sec:problem}), the planner can still see occupants at the step after the trim and switch the AHUs back on; v1x ignores the count for half an hour after the evening hour (Appendix~\ref{app:later}).

\subsection{Plant mapping}\label{app:plantmap}
The plant layer converts each zone's interval to commands in closed form with the simulator's device equations (Appendix~\ref{app:plant}).
Each zone gets the lowest airflow that delivers its heat: a cooling zone the flow that removes exactly $-q^{\mathrm{hi}}_i$, a heating zone the flow at which air discharged at the 308.15\,K VAV limit delivers $q^{\mathrm{lo}}_i$.
The AHU's duct static pressure is the lowest that lets its neediest zone reach that flow, and the dampers follow.
Where the supply air alone delivers less than $q^{\mathrm{lo}}_i$, the reheat valve opens just enough to reach the required discharge temperature; the pump head is set by the zone that needs the most hot water, and the pump runs only while some zone reheats.
Without cooling demand, an AHU runs at its maximum supply-air setpoint (305\,K) and 70\% recirculation; with cooling demand, it picks the setpoint in 285--293\,K, with economiser recirculation, that minimises an estimate of fan, chiller and reheat power \eqref{eq:satcost}.

\subsection{Constants}
\label{sec:constants}
Ten constants (band margins, $H$, $\tau_d$, $\kappa$, $\ell$, the holiday rule and three night-flush thresholds) are the mean of the fourth CMA-ES generation in a search on short windows of 12 dev30p scenarios, chosen among the generation means on 12 full-length ROM selection scenarios (10 from dev30p, 2 from a broader generated set); Appendix~\ref{app:details} lists them unrounded, with the hand-set values they replaced.
The same search also made RLS forget faster ($\lambda = 0.99906$, a memory of about 1{,}060 instead of 2{,}000 steps) and lengthened the slab constant.
On the hard18x set, this change made one extreme-cold scenario score 1.49 times as high (worse), and reverting this pair (and no other group of constants) restored it, so $\lambda = 0.9995$ and the 24-step slab constant are kept.
We did not separate the two constants of the pair.
hard18x therefore caused one revert, on top of its part in the selection rule (Appendix~\ref{app:selection}), and we report it as a stress set rather than an independent test (\S\ref{sec:results}).

\subsection{Selection history and calibration}\label{app:selection}
\paragraph{Selection history.}
Constants were tuned on dev30p and the reduced-model selection scenarios.
val18p decided which versions we kept, and from v1n on hard18x did too: a candidate had to improve val18p and hard18x pooled by at least 1\%, and from v1q (the tuned constants with the two RLS constants reverted) on to have paired 95\% intervals below 1 on val18p (both simulators) and on hard18x; under both rules the cold hard18x scenarios together could not get more than 1\% worse.
hard18x also caused one revert: it showed that the tuned RLS constants failed in one extreme-cold scenario, and we reverted them (\S\ref{sec:constants}).
Neither set is therefore a clean held-out test, and we report hard18x as a stress set.

\begin{figure}[h]
  \centering
  \includegraphics[width=\linewidth]{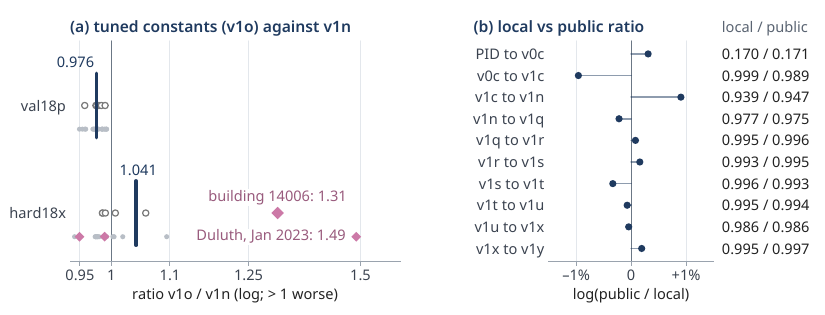}
  \caption{(a) v1o (v1n with the CMA-ES constants) against v1n (formal gates): bar, all 18 scenarios; circles, buildings; dots, scenarios; diamonds, building\_14006 and its scenarios. On val18p every scenario improves; on hard18x the aggregate is 4.1\% worse, and building\_14006 accounts for all of the net loss (101\%), its Duluth January scenario alone for 103\% (the shares exceed 100\% because the other five buildings net a small gain: three improve, 14014 is 5.7\% and 14000 0.6\% worse). (b) Ratio of the public-leaderboard scores of consecutive submitted versions against the ratio of their local scores (val18p and hard18x pooled), with both ratios printed at the right; the largest log error is 0.0096.}
  \label{fig:stress}
\end{figure}

\paragraph{Calibration against the leaderboard.}
For ten pairs of consecutive versions that we submitted, the ratio on our local sets (the 36 val18p and hard18x scenario scores pooled) predicted the ratio on the public leaderboard with a log error of at most 0.0096, and at most 0.0033 for the last six, from v1q to v1y, as Figure~\ref{fig:stress}(b) shows; public scores are rounded to five decimals, which alone allows a log error of about 0.001.

\subsection{Later versions}
\label{app:later}
After v1r we kept the model and the planner and added eleven refinements at schedule edges and plant limits, none of which uses the billing gap; the latest, v1y, scores 0.973, 0.964 and 0.965 times v1r on val18p, hard18x and the public leaderboard.
v1s heats the cold mixed air of zones held at their damper minimum on heating-only AHUs to 0.5\,K above the target band's lower edge $L$ when the mean outside air is below $-5$\,\textdegree C.
v1t keeps every zone above $L - 8.5$\,K in the 36 steps before a start when the outside air is below $-20$\,\textdegree C.
v1u narrows the cooling margin from 0.26 to 0.12\,K after 09:00.
v1w adds four: the hot-water setpoint drops to 310\,K once the pump has been off for 24\,h in weather above 15\,\textdegree C and returns to its maximum at the step the pump restarts, so it is at its maximum whenever reheat can draw on it; supply-air setpoints are scanned from 285 to 305\,K in 1\,K steps; the hourly heating margin is learned from where the zones actually sit, never above v1u's; and margins fall by at most 0.02\,K per step.
v1x ignores the occupant count for half an hour after the evening hour (Appendix~\ref{app:flush}).
v1y adds three: once the building has held at least 12 occupants, the wait before a weekday counts as a holiday shrinks so that, under the simulator's arrival rate of 1/36 per occupant and step, a working day is taken for a holiday with probability at most 0.0002 (never longer than the tuned 0.979\,h, and back to it for the rest of the episode once a working day's first arrival comes later than the shortened wait); optimal start heats at $\kappa = 0.752$ of the modelled capacity, the share its test already assumes, instead of all of it; and the night-flush test for a cooling day reads the last occupied day rather than the calendar day before, so an early holiday call no longer cancels the next morning's flush.
Each main version passed our selection rule against its predecessor (paired 95\% upper bounds below 1 on val18p stock and boiler-corrected and on hard18x stock, and at most 1\% worse on the seven coldest hard18x scenarios); v1t missed it on val18p (upper bound 1.0004) and was submitted only as a calibration point, and v1u, which includes it, passed against v1s.
Table~\ref{tab:later} lists the scores.

\begin{table}[h]
  \centering
  \caption{Later versions (formal gates; public: Kaggle public leaderboard). v1w was not submitted on its own.}
  \label{tab:later}
  \small
  \begin{tabular}{lcccc}
    \toprule
    policy & val18p stock & val18p boilerfix & hard18x stock & public \\
    \midrule
    v1r (analysed in the paper) & 0.004956 & 0.004956 & 0.016088 & 0.00996 \\
    v1s & 0.004945 & 0.004945 & 0.015959 & 0.00991 \\
    v1t & 0.004937 & 0.004937 & 0.015889 & 0.00984 \\
    v1u & 0.004889 & 0.004889 & 0.015825 & 0.00978 \\
    v1w & 0.004859 & 0.004836 & 0.015692 & -- \\
    v1x & 0.004833 & 0.004811 & 0.015593 & 0.00964 \\
    v1y & 0.004821 & 0.004798 & 0.015503 & 0.00961 \\
    \bottomrule
  \end{tabular}
\end{table}

\subsection{Time course of the ablation}\label{app:time}
\begin{figure}[h]
  \centering
  \includegraphics[width=\linewidth]{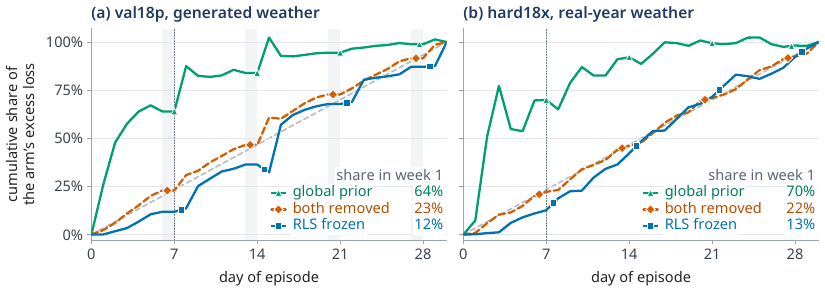}
  \caption{When each ablated part loses. For each arm, the cumulative share of its 30-day excess loss over v1r (summed over the 18 scenarios of each set, reduced model), by day of the episode; the dashed diagonal is an even spread, the dotted line marks day 7, and the numbers are the shares lost by then; markers sit on their lines at weekly intervals; a share can pass 100\% for a few days where the arm does better than v1r. Light bands on val18p: the four days on which v1r has almost no loss (unoccupied). The prior's loss is front-loaded and returns after the first two unoccupied days; identification's loss accrues over the weeks.}
  \label{fig:time}
\end{figure}

Figure~\ref{fig:time} shows, for each arm of the ablation in \S\ref{sec:results}, when its excess loss over v1r accrues.
In the numbers of Figure~\ref{fig:ablation}, freezing the RLS update costs 4.4\% on val18p and 8.1\% on hard18x, and replacing the zone-map prior by the global prior while keeping RLS 2.3\% and 1.7\%; replacing only the heat-gain part of the prior costs 2.9\% on val18p and 1.7\% on hard18x, and only the envelope part 0.3\% on both. On the hold18 holdout the five arms cost 1.9\%, 2.8\%, 2.5\%, 0.4\% and 42.1\%. With one building left out, the five ratios of Figure~\ref{fig:ablation} range over 1.040--1.049, 1.019--1.027, 1.021--1.034, 1.002--1.004 and 1.175--1.330 on val18p; 1.066--1.086, 1.008--1.022, 1.006--1.025, 1.002--1.003 and 1.133--1.256 on hard18x; and 1.014--1.023, 1.025--1.030, 1.022--1.030, 1.002--1.004 and 1.367--1.537 on hold18.
Replacing the floor-plan prior loses 64\% (val18p) and 70\% (hard18x) of its excess in the first seven of the 30 days, and on val18p again on the first day after each of the first two unoccupied days, when the building is heated from cold.
Freezing identification loses only 12\% and 13\% of its excess in the first week, less than an even spread (23\%), and with both removed the first-week share is close to an even spread (23\% and 22\%).

\subsection{Interaction per building}\label{app:interaction}
Every building is worse in every arm, except one stress-set building that is unchanged (0.9999) when only the envelope part of the prior is replaced and one holdout building that is 0.5\% better with RLS frozen.
For each building, with $r$ the loss ratio to v1r of an arm over the building's three scenarios (ROM), the interaction $r_{\mathrm{both}}/(r_{\mathrm{frozen}}\,r_{\mathrm{global}})$ is 1.00--1.59 on val18p (above 1 in all six buildings), 0.99--1.36 on hard18x (above 1 in five) and 1.06--1.71 on hold18 (above 1 in all six); it is largest in buildings 18017 (1.71), 10009 (1.59), 18015 (1.54) and 18009 (1.52).
Over all six buildings it is 1.20, 1.09 and 1.36, and 1.10--1.25, 1.04--1.15 and 1.31--1.46 when one building is left out.
It is a no-interaction reference on the multiplicative scale, not a test of statistical independence.
Comfort regret carries 62\% (val18p), 50\% (hard18x) and 72\% (hold18) of the excess loss of the arm with both removed, and 13--94\% per building.

\subsection{Where the loss is}
Almost all of the remaining score is energy and carbon.
For v1n, comfort regret is 1.5\% of $|\bar r|$ on val18p and 2.2\% on hard18x; energy and carbon share the rest about equally.
Heating fuel in cold scenarios dominates both, so the levers that matter act on how much heat the plant delivers and how efficiently.

\subsection{RLS safeguards and the observation lag}\label{app:lag}
$P$ is rescaled to unit trace when its trace exceeds 1, the first four coefficients are projected onto physical ranges (Appendix~\ref{app:details}), and steps with a non-finite or $\ge 3$\,K increment are skipped.

\emph{Observation lag.} Observed zone temperatures lag the plant by one step (\S\ref{sec:problem}), so the transition $y_{k-1} \to y_k$ is paired with the command issued at step $k-2$.
To act on the current state, the policy advances $y_k$ one model step with the command issued at $k-1$ and adds the disturbance estimate $d$ of \S\ref{sec:dist}; this estimate $\hat{x}_k$ and $m_k$ start all planning.

\subsection{Room labels}\label{app:rooms}
The policy labels the rooms of the zone map with \texttt{scipy.ndimage.label}, matches \texttt{room\_$N$} to VAV box $N$, and counts each room's cells $n_i$ and exterior-contact cells $e_i$ (room cells within two 4-neighbour steps of exterior space).

\subsection{Scale of the heat-gain prior}\label{app:scale}
With $s = 1$, the prior overestimated the heat gain in warm scenarios; RLS took weeks to bring it down, and in the meantime the planner expected more cooling per kW than it got.
In a reduced-model run (relative tolerance $10^{-4}$, boiler-corrected simulator) of the hot development scenario 12004/phoenix\_4 (June), the zones' median heat gain started at 0.377, 3.3 times the global prior, was still 0.274 after two weeks and 0.181 after three, and ended at 0.111, within 0.001 of every other setting; with $s = 0.35$ it started at 0.132.
The scale is therefore set once from the first outside-air reading, $s = 1 - 0.65\,[(T^{\mathrm{o}}_0 - 278\,\mathrm{K})/15\,\mathrm{K}]_0^1$: the full prior at or below 278\,K, 0.35 of it at or above 293\,K.

\subsection{Occupancy rules}\label{app:occ}
A step is \emph{occupied} if it falls on a weekday within the scheduled hours or the occupant count is positive; a weekday with no occupants between 0.98\,h after the morning start and noon is treated as a holiday, and no occupants at or after 13:00 ends the day.

\subsection{Capacity limits of the interval}
The interval is limited to the zone's heating and cooling capacity $\bar{Q}^{\mathrm{h}}_i, \bar{Q}^{\mathrm{c}}_i$, computed from the plant equations at the current state (Appendix~\ref{app:plant}), and the heating end wins if the ends conflict.
An AHU runs only if one of its zones needs action ($q^{\mathrm{lo}}_i > 0$ or $q^{\mathrm{hi}}_i < 0$).

\subsection{Optimal start}\label{app:start}
$S_k$ is the number of steps until the next weekday morning start.
When the next start is under 36\,h away, every third step the model is rolled forward $n = \min(S_k, 192)$ steps with the current outside temperature and $c = 0$, once free ($x^{0}_n$) and once with 1\,kW from step $\ell = 2$ on, giving the unit response $u_i = x^{1}_{n,i} - x^{0}_{n,i}$.
A zone heats at full capacity now ($q^{\mathrm{lo}}_i = \bar{Q}^{\mathrm{h}}_i$) if $x^{0}_{n,i} + \kappa\, u_i\, \bar{Q}^{\mathrm{h}}_i < L$ with $\kappa = 0.75$, that is, if starting $\ell$ steps later with 75\% of the modelled capacity would miss the band; cooling is the mirror image, and heating wins if both hold.
The decision is not latched, so a zone conditions only while a later start would miss the band; within $\ell$ steps of the start, \eqref{eq:interval} applies.

\subsection{Observations}\label{app:obs}
Observations are the zone air temperatures, setpoints, damper and valve positions and airflow setpoints of every VAV box; supply temperature, static pressure, airflow and recirculation of every air handler (AHU); the hot-water supply temperature; the outside air temperature; the number of occupants in the whole building; and the time.

\subsection{Time-of-day weights in the score}
\label{app:tou}
The reward divides each step's energy cost, and separately its carbon, by those of the configured maximum draw (80\,kW of electricity and 205\,kW of gas per AHU) priced at the same moment (\texttt{setpoint\_energy\_carbon\_regret.py} in the starter kit).
With the default tables the weekday electricity price varies by a factor of 1.25 over the day, the weekend price not at all, and grid carbon by 1.19.
Over every day of a year, the resulting score weight on a unit of electricity varies by at most 8.7\% within the day and on gas by at most 6.6\%, and the ratio of the two stays between 1.81 and 2.35.

\subsection{What the score does not see: summary table}\label{app:unseen}
Table~\ref{tab:unseen} collects what \S\ref{sec:analysis} measures outside the score.
Correcting the boiler billing leaves the submitted policy's score unchanged (0.004956 on the validation set in both simulators; the hand-set version scores 0.016532 and 0.016531 on the stress set, where cold scenarios reheat most) and makes the PID baseline's 5.6\% worse; as a diagnostic only, an earlier variant of ours whose setpoints use the billing gap scores 22.1\% better on the validation set (0.005415 to 0.004220), and we never submitted it.
The return-air imbalance moves 0.19\% of boiler heat for our policy against 2.8\% for the baseline.
In the coldest scenario our policy lets empty zones fall far below freezing with 38\% of the baseline's boiler heat, and the last rows give the price of a soft frost target.

\begin{table}[h]
  \centering
  \caption{What the score does not see. Billing: formal gate on val18p; all other rows reduced model (ROM). The coldest scenario is hard18x building\_14008 in Fairbanks, January 2022; the frost rows pool the six cold hard18x scenarios.}
  \label{tab:unseen}
  \small
  \begin{tabular}{lll}
    \toprule
    quantity & v1r & PID baseline \\
    \midrule
    score change, billing-corrected simulator & 0.0\% & $+5.6\%$ \\
    \addlinespace
    return-air heat to zones / boiler heat & $+0.19\%$ (36 runs) & $+2.85\%$, $+2.83\%$ (2 runs) \\
    \midrule
    \multicolumn{3}{l}{\emph{coldest scenario, no frost floor}} \\
    \quad zone-steps below 5\,\textdegree C / below 0\,\textdegree C & 27\% / 19\% & 0.09\% / 0\% \\
    \quad coldest zone; coldest zone while occupied & $-30.3$ / 9.8\,\textdegree C & 2.2 / 7.1\,\textdegree C \\
    \quad boiler heat & 149\,MWh & 396\,MWh \\
    \midrule
    \multicolumn{3}{l}{\emph{soft frost target added to v1r (six cold scenarios)}} \\
    \quad 5\,\textdegree C: score; boiler heat & $\times 1.077$; $\times 1.110$ & -- \\
    \quad 10\,\textdegree C: score; boiler heat & $\times 1.145$; $\times 1.190$ & -- \\
    \quad 5\,\textdegree C, coldest scenario: below 5\,\textdegree C; coldest zone & 0.8\%; 1.6\,\textdegree C & -- \\
    \bottomrule
  \end{tabular}
\end{table}

\subsection{Return-air mixing}\label{app:return}
We recorded the energy balance proposed in starter-kit \href{https://github.com/smart-buildings-2026/starter-kit/pull/3}{PR~\#3} (by StnCsn) at every step of every reduced-model run: the heat that the area-weighted return temperature adds or removes relative to the airflow-weighted mixture of the air actually extracted, counted where it reaches the zones (ROM).
Over 36 full-length runs of v1r (all of val18p and hard18x) it nets $+0.19\%$ of the boiler heat, between $-2.1\%$ and $+2.5\%$ per scenario and within $\pm 0.9\%$ in the ten scenarios that need the most heat.
The PID baseline receives $+2.85\%$ and $+2.83\%$ in the two cold scenarios where we ran it, 9 to 17 times v1r's absolute amount.
By this measure v1r does not gain from the imbalance to any material degree; we did not measure how its score would change on a corrected simulator.

\subsection{Frost target: further numbers}\label{app:frost}
\begin{figure}[!tbh]
  \centering
  \includegraphics[width=\linewidth]{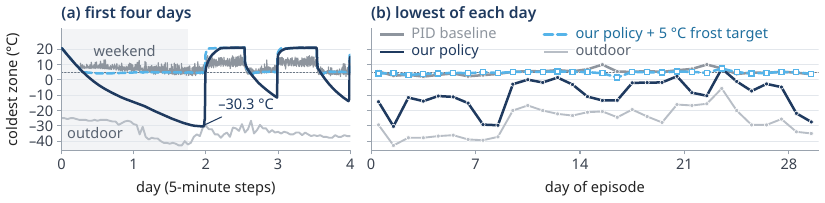}
  \caption{The coldest stress-set scenario (14008, Fairbanks, Jan.\ 2022; ROM): coldest-zone temperature per controller. (a) Days 1--4, from a Saturday; (b) daily minimum (markers). Dotted: 5\,\textdegree C.}
  \label{fig:cold}
\end{figure}

\begin{figure}[h]
  \centering
  \includegraphics[width=\linewidth]{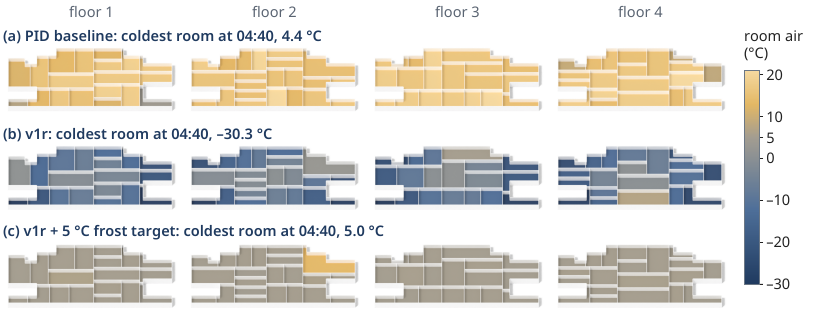}
  \caption{The building of Figure~\ref{fig:cold} at the coldest moment of v1r's episode, Monday 3 January 2022, 04:40, before the morning preheat: every room of its four floors coloured by its air temperature (reduced model). Rows: (a) the PID baseline, (b) v1r (the submitted policy, ``our policy'' in Figure~\ref{fig:cold}), (c) v1r with the 5\,\textdegree C frost target. Nobody is in the building at this hour, so none of the three is charged any comfort regret; only the energy and carbon of the heat they use count.}
  \label{fig:coldnight}
\end{figure}

Figure~\ref{fig:cold} follows the coldest zone of the coldest scenario under each controller: under our policy it drops over the weekend and recovers for each working day, while the PID baseline holds it above freezing; Figure~\ref{fig:coldnight} shows every room at the coldest moment.
The 5\,\textdegree C target's cost on the six cold stress-set scenarios (the December and January strips of Table~\ref{tab:sets} except Chicago's) is 1.02--1.19 per scenario.
The target is an extra lower end of each zone's heat interval, aimed 0.5\,K above the stated temperature.

On the six val18p scenarios with cold-month weather, the corresponding costs are factors of 1.035 and 1.085 (ROM).
It is a planning target, not a guarantee: with the 5\,\textdegree C target, 0.8\% of zone-steps in the coldest scenario stay below 5\,\textdegree C and the minimum is 1.6\,\textdegree C.
The cost also buys comfort: the coldest occupied zone, at 9.8\,\textdegree C in the coldest scenario without the target, stays at or above 16.3\,\textdegree C in all six with it.
Part of the cost is the simulator's plant: every AHU recirculates at most 70\%, so heating an empty building also heats at least 30\% outside air.

\subsection{Gate}\label{app:gate}
A formal score runs every scenario of a set to the end of the episode in the official simulator, with the same policy file that is submitted.
On the public example scenario our gate gives a mean reward of $-0.024907$ for the PID baseline, where the Kaggle gateway reports $-0.0249$ (a score of 0.0249).
Every candidate is scored on the stock simulator and on the boiler-corrected copy; a policy that gains from the billing gap shows up as a difference between the two.
The random mixing and the occupancy model use fixed seeds, so with the same code and libraries a rerun gives the same score bit for bit.
On two dev30p scenarios truncated to 2{,}000 steps, changing the mixing seed changed v1n's score by less than 0.0001\%, and 0.01\,K white noise on every zone-temperature observation by at most 0.041\%; those checks did not redraw the occupancy seed (the val18p redraw is in Appendix~\ref{app:comparisons}).

\subsection{Reduced-order model}\label{app:rom}
For screening and for the ablations we replace the simulator's thermal-grid solve by a Galerkin reduced-order model of the same grid and drop its random mixing within rooms; the devices and the reward are the official code.
The reduced model projects the grid's backward-Euler system onto an orthonormal basis that spans the grid's responses to every input (the heat of each zone, the air-side terms and the ambient temperature) at six timescales from one step to one week (1, 3, 12, 48, 288 and 2{,}016 steps), with two Krylov steps at the shortest.
Directions are added by block Gram--Schmidt and kept if their singular value exceeds $10^{-6}\sqrt{m}$ for $m$ unit-norm responses, the relative tolerance of $10^{-6}$.
The thermal-mass nodes get their own basis, the same basis restricted to them, and the air-side coupling is solved exactly in the reduced space (Woodbury identity).
Each building's model is built once from the simulator's own matrices; it needs simulator internals, so a submitted policy cannot use it.
The official scenarios also swap control volumes at random within each room as a stand-in for convection; the reduced model switches this off, and zone means come from the reduced state.
With both approximations, our policy's scenario scores differ from the formal gate by at most 0.0075\% on the 36 scenarios where we have both (\S\ref{sec:eval}).
We also reran the coldest scenario of \S\ref{sec:analysis} with the official solver, for v1r and for the 5\,\textdegree C frost target: scores and boiler heat agree within 0.003\%, the coldest zone and the coldest occupied zone within 0.001\,K, and the shares of zone-steps below 5\,\textdegree C and 0\,\textdegree C within 0.002 percentage points, confirming these two runs' temperature results on the official solver.

\subsection{Comparisons}\label{app:comparisons}
The submitted policy also beats our earlier versions in all six buildings of every comparison (0.924 times v0c on the stress set, 0.969 and 0.973 times v1n).
Two policies on the same set are compared by the ratio $\sum_s |\bar r^{\mathrm{cand}}_s| / \sum_s |\bar r^{\mathrm{ref}}_s|$, which is the ratio of their scores.
Scenarios that share a building are not independent, so the primary evidence is per building: how many of the six buildings improve (6 of 6 has a two-sided sign-test $p = 0.031$) and the range of the ratio when one building is left out.
Against v0c on hard18x the leave-one-building-out range is 0.914--0.927. For the comparison with v1n we also give paired 95\% bootstrap intervals (10{,}000 resamples): over buildings 0.964--0.975 on val18p and 0.970--0.975 on hard18x, over scenarios 0.964--0.974 and 0.969--0.977; with six buildings the former are descriptive. With the occupancy of val18p redrawn (seed 20261109 instead of 17321), v1r scores 0.004949 (0.004956 before) and v1n 0.005111 (0.005115); the ratio is 0.968, with v1r better in 18 of 18 scenarios and intervals 0.963--0.975 over buildings and 0.963--0.974 over scenarios.
The holdout of \S\ref{sec:results} was generated on 9 November with building seeds from 18000 and NOAA ISD-Lite observations from 2023 at the Boston, Seattle, Kansas City, Salt Lake City, Nashville and Sacramento airports, in strips starting on 10 January, 10 July and 1 October; building $i$ takes station $i$ in January, station $i{+}1$ in July and station $i{+}2$ in October. Its design was committed before any run. v1r scores 0.004004 and the PID baseline 0.042057 (formal gates, stock simulator); the 95\% bootstrap intervals of their ratio (10{,}000 resamples, seed 0) are 0.075--0.119 over buildings and 0.076--0.118 over scenarios.

\subsection{Guideline 36 baselines}\label{app:g36}
Both rule-based baselines were written for this benchmark by C.F.\ as single numpy files with no weights and no training; they read all indices from the building metadata and never the building identifier, weather name or path.
The basic version (G36, public score 0.01385, submitted 15 October) has five blocks: per-zone velocity PI loops with a dead band inside the comfort band; a VAV dual-maximum mapping that opens the reheat valve before raising airflow; trim-and-respond of the hot-water temperature with an outdoor-air reset floor; optimal start with an unoccupied setback to 285 and 301\,K; and trim-and-respond of the supply-air temperature and static pressure with an economizer.
The tuned version (G36-t, the team's competition entry, public score 0.01044) keeps these blocks and adds start-ramp learning from a quantile of the miss, holiday detection, trim-and-respond of the pump pressure, cold- and mild-climate corrections weighted by the observed 24-hour mean outdoor temperature, a hot-water flow cap, and a deep-cold coast that lowers the unoccupied floor to 270\,K. For the frost-matched row of Table~\ref{tab:main} this floor is 278.15\,K (5\,\textdegree C) and nothing else changes.
Its constants come from CMA-ES searches on short development windows; two of its steps were adopted because they improved the public score.
Formal gate scores, stock simulator: G36 0.007430, 0.022678 and 0.006700 on val18p, hard18x and the holdout; G36-t 0.005513, 0.016779 and 0.004805 on val18p, hard18x and the holdout, and 0.005493 on val18p with the billing corrected.
The submitted policy is better than G36 in all 54 scenarios and than G36-t in 53, in all six buildings of each set against both; the one loss is building 14014 in Bismarck in December 2022 (hard18x), 1.010 times G36-t.
With the 5\,\textdegree C frost target the submitted policy is worse than G36-t in four cold stress-set scenarios and two of the six stress-set buildings; against the frost-matched G36-t, five cold stress-set scenarios (at most 1.032) and one stress-set building (1.002) are worse, and no scenario on the other two sets.

\subsection{Second simulator: BOPTEST}\label{app:boptest}
We ran BOPTEST v0.9.0 (release commit 9b1610bf) case \texttt{multizone\_office\_simple\_air} locally, calling the framework's own test-case class in a single container; simulation, scenarios and indicators are BOPTEST's code, unmodified.
Each run warms up for one week under the baseline controller and then tests for two weeks at 5-minute steps (4032 steps), with the constant electricity price.
The prior uses the case's geometry: floor areas of 985\,m$^2$ (core), 208\,m$^2$ (north, south) and 131\,m$^2$ (east, west), and exterior wall lengths of 49.9 and 33.3\,m from the model's wall-plus-window areas, converted to the competition's 0.2\,m cells.
The competition's constants (1000, 1.1, bounds, $P_0$) are unchanged; with them the perimeter zones' envelope coefficient starts at its cap of 0.05.

The thermal model, prior, RLS, disturbance estimate and planner are the submitted policy's code.
The mapping to the plant differs in five places.
Observations have no lag (BOPTEST returns the state at the end of each step).
The model's heat input is computed from measured VAV airflow and discharge temperature.
Heating comes only from VAV reheat with the air-handler coils closed, as in the competition's plant; supply-air temperature in cooling is chosen by the competition's rule with all-electric energy weights.
During occupied hours each air handler runs at least at the baseline's minimum airflow, with the outdoor-air damper on a CO$_2$ loop (800\,ppm), because BOPTEST scores indoor air quality and the competition does not.
The case schedule has no weekends, so every day is a working day.
Every override was confirmed active in every run, and a repeated run gave bit-identical indicators and trajectories.
Table~\ref{tab:boptest} gives the indicators.

\begin{table}[h]
  \centering
  \small
  \caption{BOPTEST indicators per arm and scenario (two-week test periods; lower is better). Thermal discomfort in K\,h per zone, energy in kWh\,m$^{-2}$, CO$_2$ discomfort in ppm\,h per zone. Cost and emissions are proportional to energy here (all-electric plant, constant price) and are not shown. Baseline: the case's own controller. One case, so the four scenarios are not independent buildings.}
  \label{tab:boptest}
  \begin{tabular}{lcccc}
    \toprule
    & \multicolumn{2}{c}{heating} & \multicolumn{2}{c}{cooling} \\
    arm & peak & typical & peak & typical \\
    \midrule
    \multicolumn{5}{l}{\emph{thermal discomfort}} \\
    full (floor-plan prior + RLS) & 2.81 & 0.55 & 7.35 & 0.87 \\
    RLS frozen & 2.27 & 1.35 & 7.97 & 1.17 \\
    global prior & 3.68 & 0.63 & 7.70 & 0.96 \\
    both removed & 1.68 & 0.55 & 6.38 & 0.70 \\
    baseline & 11.84 & 7.70 & 8.61 & 2.67 \\
    \midrule
    \multicolumn{5}{l}{\emph{energy}} \\
    full (floor-plan prior + RLS) & 1.434 & 0.702 & 2.133 & 0.769 \\
    RLS frozen & 1.511 & 0.977 & 2.130 & 0.780 \\
    global prior & 1.413 & 0.661 & 2.136 & 0.774 \\
    both removed & 1.527 & 0.758 & 2.140 & 0.792 \\
    baseline & 1.790 & 1.052 & 2.188 & 0.848 \\
    \midrule
    \multicolumn{5}{l}{\emph{CO$_2$ discomfort}} \\
    full (floor-plan prior + RLS) & 913 & 870 & 0 & 12 \\
    RLS frozen & 2405 & 4486 & 2 & 15 \\
    global prior & 272 & 772 & 0 & 38 \\
    both removed & 1287 & 4767 & 0 & 0 \\
    baseline & 0 & 0 & 80 & 20 \\
    \bottomrule
  \end{tabular}
\end{table}

With identification on, the perimeter zones' envelope coefficient ends at 0.012--0.018 from the floor-plan prior and 0.009--0.014 from the global prior in the heating scenarios; frozen, it stays at 0.05 or 0.018.
The frozen floor-plan arm uses 2.31 and 2.57 times the full controller's heat-pump energy at night (19:00--6:00) in the peak and typical heating scenarios (989 against 428 and 640 against 249\,kWh), consistent with a model that overstates heat loss and so heats earlier.
The two frozen arms also have the most CO$_2$ discomfort in heating.
In an exploratory run fixed before the ablation results were seen and done once, we refitted the two prior constants on a baseline run outside the test periods (131.5 for 1000 and 0.046 for 1.1) and reran the four arms in the two heating scenarios: the arms nearly converge (thermal discomfort within 4--11\%, energy within 1--2.5\%), and against the same arm with the competition's constants each uses less energy (0.68 to 0.99 times) with 1.5 to 3.8 times the thermal discomfort.
We did not examine why RLS then moves the heat-gain coefficients well above the refitted start, nor why the frozen arms have less CO$_2$ discomfort in that run, and we did not combine the indicators into one ranking.

\section{Reproduction}
\label{app:repro}
\paragraph{Software.}
The rules allow sharing code during the competition only on Kaggle (\href{https://www.kaggle.com/competitions/neurips-2026-smart-buildings/rules}{Section~3.6}), so we will release the scripts of this appendix after 15 January 2027.
The policy is a single Python file that imports only the standard library, NumPy and SciPy.
Formal scores (\S\ref{sec:eval}) come from a scorer that runs a policy file the way the Kaggle gateway does: a plain \texttt{reset}/\texttt{step} loop to the end of each episode (8{,}641 rewarded steps, as the gateway ran on Kaggle; the competition page states 8{,}640), a fresh policy per scenario, metadata only at construction, and the gateway's rejection of non-finite, out-of-range or misshaped actions.
It runs the simulator that ships with the competition data under Python 3.12.3, NumPy 2.0.2, SciPy 1.16.3, Numba 0.60.0 and pandas 2.3.3; the floor-plan and scenario generators are those of the starter kit (\texttt{smart-buildings-2026/starter-kit}, commit \texttt{95cea5b}).

\paragraph{Commands.}
\texttt{scripts/verify.sh} \emph{file}\texttt{.py:}\emph{Class} \texttt{-{}-set val18p} gives a formal score on a set, and \texttt{-{}-sim boilerfix} switches to the boiler-corrected copy.
\texttt{scripts/gen\_buildings.py} draws buildings from the starter kit's floor-plan generator in ranges slightly wider than those of the six shipped buildings (width 310--440 and height 185--305 cells against 318--430 and 188--301; 2--4 storeys and smallest room 25--35 cells as shipped) and leaves every other generator setting at its default.
\texttt{scripts/make\_scenarios.py -{}-set hard18x} builds the stress set from the weather that \texttt{scripts/fetch\_weather.py} downloads and converts.

\paragraph{Weather.}
val18p uses official weather strips that the training set does not use.
hard18x uses hourly observations from NOAA's Integrated Surface Database (ISD-Lite; \citealp{noaa2026isd}) at the airport stations of 14 cities, downloaded from NOAA's open-data copy.
We convert them to the official CSV format (local standard time minus one hour, as in the official files), fill gaps of at most six hours by linear interpolation, and compute humidity from temperature and dew point with the Magnus formula.
Data from U.S. stations are U.S. government works in the public domain; NOAA marks ISD data from non-U.S. stations, here Singapore and Riyadh, as subject to WMO Resolution 40, which does not allow redistribution, so we do not redistribute those two strips.
Table~\ref{tab:sets} lists the buildings and strips of the validation and stress sets; Appendix~\ref{app:comparisons} describes the holdout.

\begin{table}[h]
  \centering
  \small
  \caption{Buildings and weather strips of the validation and stress sets: size is width $\times$ height in floor-plan cells $\times$ storeys; rooms and AHUs as generated (the scenarios of one building share them).}
  \label{tab:sets}
  \begin{tabular}{lcrrl}
    \toprule
    building & size & rooms & AHUs & weather strips \\
    \midrule
    \multicolumn{5}{l}{\emph{val18p} (287 rooms)} \\
    10001 & 401$\times$193$\times$2 & 33 & 5 & atlanta\_7, denver\_4, fairbanks\_3 \\
    10002 & 409$\times$214$\times$2 & 29 & 7 & minneapolis\_6, phoenix\_7, singapore\_5 \\
    10004 & 433$\times$266$\times$2 & 47 & 6 & atlanta\_4, denver\_1, fairbanks\_6 \\
    10005 & 425$\times$196$\times$3 & 59 & 6 & minneapolis\_4, phoenix\_2, singapore\_2 \\
    10006 & 370$\times$192$\times$2 & 24 & 3 & atlanta\_7, denver\_4, fairbanks\_3 \\
    10009 & 391$\times$211$\times$3 & 95 & 10 & minneapolis\_1, phoenix\_7, singapore\_5 \\
    \midrule
    \multicolumn{5}{l}{\emph{hard18x} (364 rooms)} \\
    14000 & 344$\times$295$\times$2 & 64 & 7 & chicago\_2019jan, el\_paso\_2024jun, miami\_2023jul \\
    14006 & 366$\times$220$\times$3 & 58 & 6 & denver\_2022oct, duluth\_2023jan, houston\_2023aug \\
    14008 & 354$\times$199$\times$4 & 80 & 8 & fairbanks\_2022jan, phoenix\_2022jul, singapore\_2024jul \\
    14014 & 369$\times$216$\times$3 & 39 & 7 & bismarck\_2022dec, chicago\_2023apr, tucson\_2023jun \\
    14036 & 328$\times$185$\times$4 & 64 & 7 & las\_vegas\_2023jul, minneapolis\_2024jan, riyadh\_2023jul \\
    14064 & 332$\times$212$\times$4 & 59 & 6 & denver\_2022dec, fairbanks\_2023dec, phoenix\_2024aug \\
    \bottomrule
  \end{tabular}
\end{table}

\end{document}